\documentstyle[aps,prl,epsf,twocolumn,floats]{revtex}
\begin{document}
\draft

\twocolumn[\hsize\textwidth\columnwidth\hsize\csname @twocolumnfalse\endcsname

\title{\bf Behavior of magnetic impurities in gapless Fermi systems}
\author{Kevin Ingersent\cite{byline}}
\address{Department of Physics, University of Florida,
P.\ O.\ Box 118440, Gainesville, Florida 32611--8440, USA}
\date{Revised version: 13 May 1996}
\maketitle

\begin{abstract}
In a number of systems, including certain semiconductors and unconventional
superconductors, the effective density of states varies near the Fermi energy
like $|E-E_F|^r$.
The behavior of dilute magnetic impurities in such systems is studied
using a non-perturbative renormalization-group approach.
Close to particle-hole symmetry, the Kondo effect is suppressed for the
cases of greatest relevance ($r=1$ and $2$).
Away from this symmetry, any quenching of the impurity moment is accompanied
by a low-temperature decrease in the impurity resistivity, rather than the
increase found in metals.
\end{abstract}

\pacs{75.20.Hr, 74.70.Tx, 72.15.Qm}

]
\narrowtext

In an interesting class of ``gapless'' Fermi systems, the density of states
$\rho(\epsilon)$ vanishes right at the Fermi energy $E_F$,
but not at any other energy near $\epsilon\equiv E-E_F=0$.
For instance, the quasiparticle density of states in an unconventional
superconductor can vary like $|\epsilon|$ or $|\epsilon|^2$ near line or point
nodes in the gap \cite{Sigrist}.
Heavy-fermion and cuprate superconductors are candidates for this behavior.
The valence and conduction bands of certain semiconductors --- including
Pb$_{1-x}$Sn$_x$Te at a critical composition \cite{Hohler}, and PbTe-SnTe
heterojunctions \cite{Volkov} --- touch in such a way that, for small
$|\epsilon|$, $\rho(\epsilon)$ is proportional to $|\epsilon|^{d-1}$
in $d$ spatial dimensions.
Electrons in a strong magnetic field (at least in the absence of disorder)
\cite{Fisher} and exotic phases of the Hubbard model \cite{Baskaran}
are also predicted to exhibit a linear pseudo-gap in two dimensions.

Antiferromagnetic coupling between magnetic impurities and a metallic
conduction band leads to a low-temperature increase in the resistivity
and a reduction in the Curie term in the susceptibility.
This Kondo effect depends on the existence of electronic excitations down to
zero energy, and thus cannot be fully realized in systems with a finite
energy gap.
Gapless systems with a density of states varying like $|\epsilon|^r$
constitute a marginal case, first studied in Ref.~\onlinecite{Withoff}.
Poor-man's scaling for the spin-\mbox{$\frac{1}{2}$}\ (impurity degeneracy
$N=2$) Kondo model, and a large-$N$ treatment restricted to $r<\frac{1}{2}$
(but recently extended \cite{Cassanello} to include $r=1$), both showed that
a Kondo effect takes place only if the electron-impurity exchange $J$ exceeds
a critical value, $J_c\propto r$; otherwise, the impurity decouples from the
band.
A large-$N$ treatment of magnetic impurities in gapless superconductors
\cite{Borkowski} yielded similar results, except that for $r\le 1$ or $N=2$,
any finite impurity concentration was found to drive $J_c$ to zero.
Recently, however, third-order scaling was applied to the $n$-channel Kondo
model to show that, at least for $n\gg 1$,
no Kondo effect can occur if $r>1/(2n)$ \cite{scaling}.

This paper reports the results of a nonperturbative renormalization-group (RG)
study of a spin-$\frac{1}{2}$ impurity interacting with an electronic band in
which $\rho(\epsilon)$ takes one of several functional forms, each
varying like $|\epsilon|^r$ near $\epsilon=0$.
A stability analysis of the RG fixed points and numerical calculations of
impurity thermodynamic properties are presented.
Particle-hole asymmetry is identified as a key factor in determining the
low-temperature physics.
At small asymmetry, the critical coupling $J_c$ above which the impurity
moment is screened becomes so large for all $r>\frac{1}{2}$ that the Kondo
effect is suppressed.
Larger asymmetries produce a low-temperature {\em decrease\/} in the impurity
resistivity, rather than the monotonic increase which characterizes the
metallic Kondo effect.

After the completion of this work, the author became aware of a similar study,
limited to particle-hole-symmetric systems and a pure power-law density of
states with $0<r\le 1$ \cite{Chen}.
The present paper confirms the conclusions of Ref.~\onlinecite{Chen} for this
special case, but shows that the generic behavior of gapless systems is quite
different.

The Kondo Hamiltonian describing impurity potential and exchange scattering of
band electrons can be written
\begin{eqnarray}
H &\equiv& H_{\text{band}} + H_{\text{imp}}
   = D \sum_{\sigma}\int\!d\varepsilon\,\varepsilon\,
        c^{\dagger}_{\varepsilon \sigma}
        c^{\rule{0pt}{1.35ex}}_{\varepsilon\sigma} \nonumber \\
  & &+ V \sum_{\sigma}\! f^{\dagger}_{0\sigma}
        f^{\rule{0pt}{1.35ex}}_{0\sigma}
     + J \sum_{\sigma,\sigma'}\! f^{\dagger}_{0\sigma}
        {\textstyle \frac{1}{2}} \bbox{\sigma}_{\!\sigma\sigma'}
        f^{\rule{0pt}{1.35ex}}_{0\sigma'} \cdot{\bf S}.
                                                        \label{H1D}
\end{eqnarray}
Here $\varepsilon=(E-E_F)/D$ is a reduced kinetic energy, measured from the
Fermi level in an isotropic band of width $2D$;
$J>0$ represents antiferromagnetic exchange;
$c^{\rule{0pt}{1.35ex}}_{\varepsilon\sigma}$ annihilates a
spin-$\sigma$ electron in an $s$-wave state of reduced energy $\varepsilon$;
and $f^{\rule{0pt}{1.35ex}}_{0\sigma} = \int\!
        d\varepsilon \sqrt{D\rho(\varepsilon D)}\;
        c^{\rule{0pt}{1.35ex}}_{\varepsilon\sigma}$
is the combination of $c^{\rule{0pt}{1.35ex}}_{\varepsilon\sigma}$'s that
destroys an electron at the impurity site.
The fermionic operators are normalized such that
$\{c^{\dagger}_{\varepsilon\sigma},
   c^{\rule{0pt}{1.35ex}}_{\varepsilon'\sigma'}\}
   =\mbox{$\delta(\varepsilon-\varepsilon')$}
   \delta_{\sigma,\sigma'}$
and
$\{f^{\dagger}_{0\sigma},f^{\rule{0pt}{1.35ex}}_{0\sigma'}\}
   =\delta_{\sigma,\sigma'}$ \cite{normalize}.

Gapless systems can be modeled by a density of states
\begin{equation}
\rho(\epsilon) \equiv \rho(\varepsilon D)= \left\{
         \begin{array}[c]{ll}
         \rho_0|\varepsilon|^r \quad & \text{if }|\varepsilon|\le 1, \\[0.5ex]
         0 & \text{otherwise},
         \end{array}
         \right.
                                        \label{puredos}
\end{equation}
with $r>0$.
This form is oversimplified, but it will turn out that for practical purposes
the results are little changed if one introduces band asymmetry,
limits the power-law variation to the vicinity of $E_F$, or
allows $\rho(0)$ to be small but nonzero.
Standard results \cite{Wilson} should (and will) be
recovered in the metallic limit, $r=0$.

Equation~(\ref{H1D}) can be treated nonperturbatively using a generalization of
Wilson's numerical RG method \cite{Wilson} to an arbitrary density of states.
The Hamiltonian is written
$
H = \lim_{M\rightarrow\infty} D\Lambda^{-M/2} \tilde{H}_M,
$
where
\begin{eqnarray}
\tilde{H}_{M} &=& \Lambda^{1/2} \tilde{H}_{M-1} + \sum_{\sigma} \left[
        \varepsilon_M f^{\dagger}_{M\sigma}f^{\rule{0pt}{1.35ex}}_{M\sigma}
                                                \right. \nonumber \\
        && \left. +\,
        \xi_{M-1}\,(f^{\dagger}_{M\sigma}f^{\rule{0pt}{1.35ex}}_{M-1,\sigma}
                + \text{h.c.}) \right]
                                                        \label{H_M}
\end{eqnarray}
for $M\ge 0$.
Here $\tilde{H}_{-1} = H_{\text{imp}}/(\Lambda^{1/2}D)$ and $\xi_{-1}=0$;
$\Lambda>1$ parameterizes the logarithmic discretization of the conduction
band \cite{Wilson};
and $f_M$ annihilates an electron in a state, centered on the impurity site,
that becomes progressively more delocalized as $M$ increases.
Numerical solutions of successive $\tilde{H}_{M}$'s capture the key physics at
a sequence of temperatures $T \approx D \Lambda^{-M/2}$ \cite{Wilson,units}.

All dependence on $\rho(\epsilon)$ is contained in $\xi_M$ and $\varepsilon_M$,
which are given by a set of Lanczos recursion relations.
Quite generally, if $\rho(\varepsilon)=\rho(-\varepsilon)$, then
$\varepsilon_M = 0$ for all $M$.
For the density of states in Eq.~(\ref{puredos}), one finds \cite{longpaper}
that
\begin{equation}
\xi_M \; \stackrel{M\rightarrow\infty}{\longrightarrow} \;
        \frac{1\!+\!r}{2\!+\!r} \;
        \frac{1\!-\!\Lambda^{-(2+r)}}{1\!-\!\Lambda^{-(1+r)}} \;
        \Lambda^{-1/2}\,\Lambda^{-(M\text{ mod }2)\,r/2} .
                                                \label{xi_M}
\end{equation}

{\em Preliminary analysis:\/}
Even before one resorts to heavy computation, study of the limits $J=0$
and $\infty$ provides valuable insight concerning the density
of states given by Eq.~(\ref{puredos}):
(1) For all $r>0$, there is a weak-coupling regime in which the Kondo
interaction is irrelevant \cite{Withoff}.
(2) At particle-hole symmetry, a Kondo effect (i.e., scaling to infinite $J$)
is ruled out for all $r>\frac{1}{2}$ \cite{scaling,Chen}.
(3) If $V\not=0$ in Eq.~(\ref{H1D}), there is a stable strong-coupling limit
in which the impurity is completely screened, but the impurity resistivity
is zero rather than taking its maximum possible value, as it does in
metals.
(4) Since each RG trajectory must flow from an unstable fixed point to a
stable one, the intermediate-coupling region is also constrained.
If the limit $J=\infty$ is stable, at least one (unstable) fixed point must
lie at some $0<J_c<\infty$, whereas instability about $J=\infty$
is consistent with uninterrupted RG flow from strong to weak coupling.
The derivation of these results, outlined in the next four paragraphs,
follows Ref.~\onlinecite{Wilson}.

For $J=0$, different electron spins decouple.
Each spin is described by an effective Hamiltonian $\tilde{H}_M^{(0)}(V)$,
where
\begin{eqnarray}
\tilde{H}_M^{(L)}(V)
   &=& \Lambda^{M/2}\, \left[\,
       (V/D)\, f^{\dagger}_{0} f^{\rule{0pt}{1.35ex}}_{0} \right. \nonumber \\
   & & + \sum_{m=L}^{M-1} \left. \Lambda^{-m/2} \xi_m
         (f^{\dagger}_{m} f^{\rule{0pt}{1.35ex}}_{m+1} + \text{h.c.}) \right] .
                                                        \label{H_M^l}
\end{eqnarray}
For $M\!\gg\!1$, it is found numerically that the low-lying eigenvalues
of $\tilde{H}_M^{(0)}(V)$ are independent of $M$ and $V$,
while the $f$'s have simple expansions in terms of the exact eigenoperators
$g_0,\ldots,g_M$:
$f_0\!=\!\Lambda^{-(1+r)M/4}\sum_j \alpha_j g_j$, where $\alpha_j$ is
independent of $M$; all other $f_{m}$'s decay with increasing $M$ at
least as fast as $f_0$ decays.
The impurity susceptibility is Curie-like, $T\chi_{\text{imp}} = \frac{1}{4}$,
and the entropy is $S_{\text{imp}}=\ln 2$ \cite{units}.
For $r>0$, electrons at the Fermi energy scatter from the impurity with a
phase shift $\delta(\epsilon=0)=0$.

Any deviation of $\tilde{H}_M$ from $\tilde{H}_M^{(0)}(V)$ must be
a combination of $f^\dagger$'s and $f$'s which respects all symmetries of
Eqs.~(\ref{H1D}) and (\ref{puredos}), multiplied by the same factor of
$\Lambda^{M/2}$ as enters Eq.~(\ref{H_M^l}).
Using the above expansion of the $f$'s and the relation
$T \approx D\Lambda^{-M/2}$, one can classify the relevance of any such
operator \cite{Wilson}.
For instance, the exchange and potential-scattering operators,
${\cal O}_J = \Lambda^{M/2}
f^\dagger_{0\sigma}\frac{1}{2}\bbox{\sigma}_{\sigma\sigma'}
f^{\rule{0pt}{1.35ex}}_{0\sigma'} \cdot {\bf S}$
and ${\cal O}_V = \Lambda^{M/2}
f^\dagger_{0\sigma} f^{\rule{0pt}{1.35ex}}_{0\sigma}$,
both vary like $T^r$, a temperature dependence which can be absorbed
into effective couplings $J_{\text{eff}}(T)=J\cdot(T/D)^r$ and
$V_{\text{eff}}(T)=V\cdot(T/D)^r$.
Any other allowed operator has an algebraically greater temperature exponent.
Thus, for all $r>0$, ${\cal O}_J$ and ${\cal O}_V$ are the leading irrelevant
operators about a stable $J=0$ fixed point.
In the metallic case ($r=0$), by contrast, $J_{\text{eff}}$ grows
logarithmically as $T$ decreases, so the fixed point is marginally unstable.

In the limit $J=\infty$ (with $V$ finite), an $f_0$ electron
is locked into a spin singlet with the impurity.
Hopping to or from $f_0$ states is ruled out, so different
electron spins again decouple.
The effective Hamiltonian is $\tilde{H}_M^{(1)}(0)$ given by Eq.~(\ref{H_M^l}).
Now $f_m\propto\Lambda^{-(1-r_1)M/4}$ for $m$ odd and
$f_m\propto\Lambda^{-(3-r_3)M/4}$ for $m$ even,
where $r_n=\min(r,n)$.
One finds that $T\chi_{\text{imp}} = r_1/8$,
$S_{\text{imp}} = 2r_1\ln 2$, and
$\delta(0)=(1-r_1)\pi/2$, indicating that even at $J=\infty$ the
impurity degree of freedom is not completely quenched, but instead
is partially transferred to the band \cite{Chen}.

The stability of this fixed point hinges on whether or not the
problem is particle-hole symmetric.
If $V=0$ in Eq.~(\ref{H1D}), the most relevant perturbation
about $\tilde{H}_M^{(1)}$ is $\Lambda^{M/2} (f^\dagger_1
f^{\rule{0pt}{1.35ex}}_1 - \frac{1}{2})^2 \propto T^{1-2r_1}$,
in which case the fixed point is stable for $r<\frac{1}{2}$,
but is destabilized for all $r>\frac{1}{2}$.
Setting $V\not=0$ admits an additional operator,
$\Lambda^{M/2}f^\dagger_{1\sigma}f^{\rule{0pt}{1.35ex}}_{1\sigma}
\propto T^{-r_1}$.
For $r>0$, this freezes the $f_1$ states as $T\rightarrow 0$
and drives the system to a new $J=\infty$ fixed point, described
by $\tilde{H}_M^{(2)}(0)$.
Due to the ``odd-even'' character of the Hamiltonians $\tilde{H}_M^{(L)}$,
$\tilde{H}_M^{(2)}(0)$ has the same low-energy spectrum and stability
properties as the $J=0$ Hamiltonian $\tilde{H}_M^{(0)}(0)$.
At this fixed point, $T\chi_{\text{imp}}=0$, $S_{\text{imp}}=0$,
and $\delta(0)=\pi$; the impurity contribution to the resistivity,
$\varrho\propto\sin^2\delta(0)$, vanishes for $r>0$.

\begin{figure}[t]
\centerline{
\vbox{\epsfxsize=87mm \epsfbox{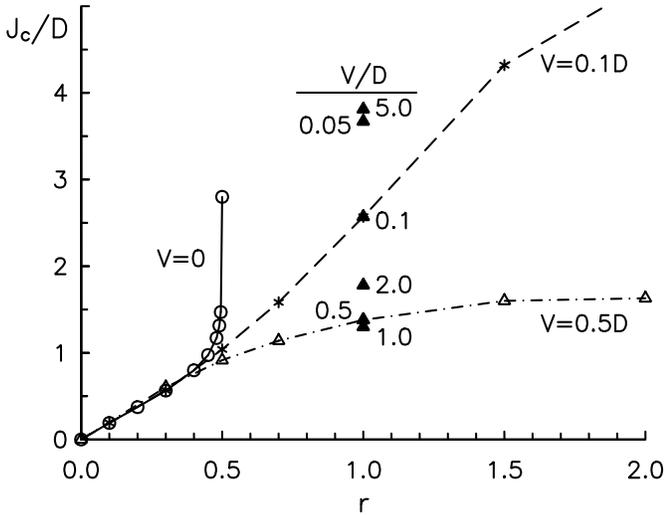}}
}
\vspace{2ex}
\caption{
Critical Kondo coupling $J_c/D$ vs. the power $r$ entering
Eq.~(\protect\ref{puredos}), for potential scatterings
$V/D=0$, $0.1$ and $0.5$.
(The connecting lines are provided as a guide to the eye.)
For $r=1$ only, data are shown for additional values of $V$.
}
\label{fig:J_c}
\end{figure}

\begin{figure}[t]
\centerline{
\vbox{\epsfxsize=87mm \epsfbox{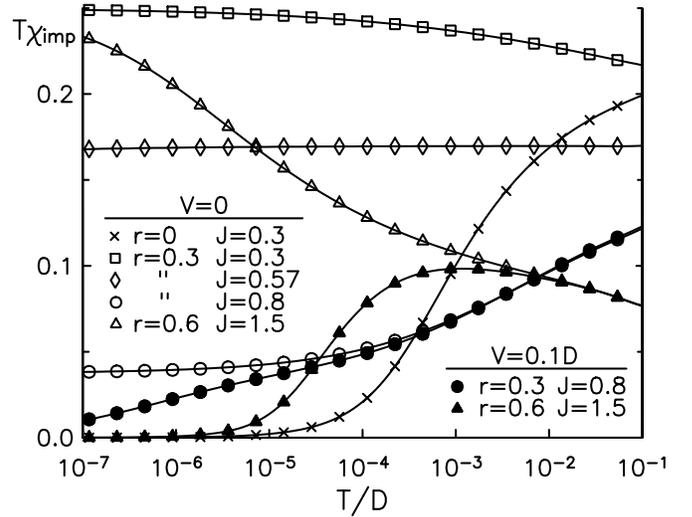}}
}
\vspace{2ex}
\caption{
Impurity susceptibility $T\chi_{\text{imp}}$ vs.\ $\log(T/D)$, for three
power-law densities of states defined in Eq.~(\protect\ref{puredos}).
}
\label{fig:chi1}
\end{figure}

\begin{figure}[t]
\centerline{
\vbox{\epsfxsize=87mm \epsfbox{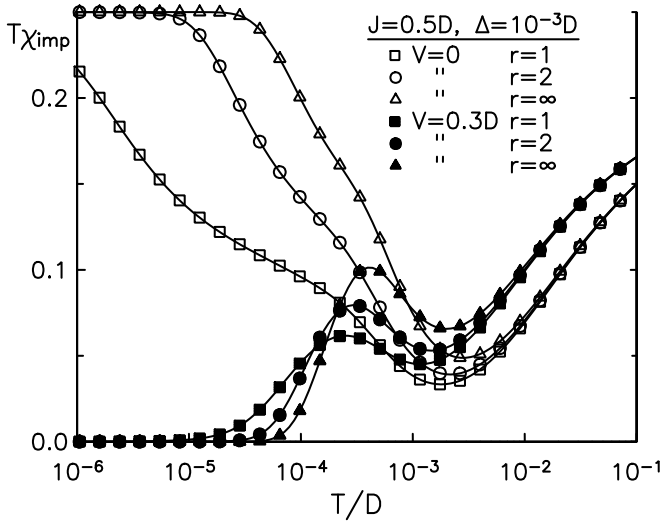}}
}
\vspace{2ex}
\caption{
Impurity susceptibility, $T\chi_{\text{imp}}$ vs.\ $\log(T/D)$,
for restricted power-law densities of states in which
$\rho(\epsilon)=$$\rho_0|\Delta/D|^r$ for $\Delta<|\epsilon|\le D$.
The $r=\infty$ curves represent an insulator.
}
\label{fig:chi2}
\end{figure}

{\em Numerical results:\/}
The RG picture has been completed by solving the Hamiltonians $\tilde{H}_M$
in Eq.~(\ref{H_M}) for arbitrary $J$.
The entropy, heat capacity and magnetic susceptibility have been computed
by adapting methods developed for the metallic Kondo problem
\cite{Wilson,Oliveira}.
(Space permits only the susceptibility to be plotted.)
The main sources of error are the discretization of the conduction band and
the truncation of the basis of $\tilde{H}_M$.
In this work 800--2000 states were retained, sufficiently many that
truncation errors are negligible (smaller than the symbols in
Figs.~\ref{fig:J_c}--\ref{fig:chi2}).
All data shown were obtained using a discretization parameter $\Lambda=3$.
Changing $\Lambda$ affects the results only through small shifts in the
value of $J_c$.

Results are presented first for pure power-law densities of states
[Eq.~(\ref{puredos})], confirming and extending the conclusions
drawn above.
Consider the particle-hole symmetric case, $V=0$.
For $0<r\le\frac{1}{2}$ there exists a critical coupling
$J_c(r)$, plotted in Fig.~\ref{fig:J_c}.
For small $r$, $J_c\approx 2rD$ (in agreement with Ref.~\onlinecite{Withoff}),
but the curve $J_c(r)$ turns upward and then terminates at $r=\frac{1}{2}$.
Far below a crossover temperature, $T_X \approx D[|J-J_c|/\max(J,J_c)]^{1/r}$,
any coupling $J\!<\!J_c$ yields the same excitation spectrum as
$\tilde{H}_M^{(0)}$, implying renormalization to zero exchange;
whereas values $J\!>\!J_c$ reproduce the strong-coupling spectrum of
$\tilde{H}_M^{(1)}$.
In the region $r>\frac{1}{2}$, the critical point disappears,
so any exchange $0<J<\infty$ renormalizes to zero \cite{scaling}.

Turning to the case $V\not=0$, one finds as expected that the curve $J_c(r,V)$
extends to arbitrary $r$ (see Fig.~\ref{fig:J_c}).
For $r\gtrsim\frac{1}{2}$, the critical coupling is strongly $V$-dependent:
As $|V|$ increases from zero, $J_c$ initially drops from infinity.
However, once the potential scattering becomes large enough to inhibit hopping
of electrons to or from $f_0$ states, further increases in $|V|$ serve only
to push $J_c$ higher.
This is illustrated by the $r=1$ data in Fig.~\ref{fig:J_c}.

Figure~\ref{fig:chi1} compares impurity susceptibilities for $r=0$, $0.3$
and $0.6$.
For $r=0$, $T\chi_{\text{imp}}$ falls monotonically to zero as the
temperature decreases --- a sign of the complete screening of the
impurity moment which occurs for any $J>0$.
This should be contrasted with the $r=0.3$ data:
For $J=0.3D$, $T\chi_{\text{imp}}$ rises at low temperatures towards
the weak-coupling value $\frac{1}{4}$; whereas the $J=0.57D$ curve is almost
flat, indicating that $J$ is very close to $J_c$.
Neither curve is greatly affected by potential scattering
(not shown in Fig.~\ref{fig:chi1}),
and in both cases $S_{\text{imp}}\approx\ln 2$ over the entire temperature
range shown.
The two $J=0.8D$ curves do exhibit impurity screening at temperatures
$T<T_X\approx 0.05D$.
For $V=0$, the susceptibility remains Curie-like, reaching a limit
$T\chi_{\text{imp}}= 0.0377$ very close to the strong-coupling value $r/8$.
The computed entropy, $S_{\text{imp}}=0.416$, is also in good agreement
with the predicted value $2r\ln 2$.
Setting $V=0.1D$ causes a second crossover, around
$T'_X=|V/D|^{1/r}T_X\approx 2\!\times\!10^{-5}D$, to the
particle-hole-asymmetric fixed point, at which
$T\chi_{\text{imp}}=0$ and $S_{\text{imp}}=0$.
Note, though, that the $V=0$ fixed point still dominates the behavior
in the range $T'_X < T < T_X$.

Figure~\ref{fig:chi1} also shows two $r=0.6$ curves.
For $V=0$, the impurity remains unscreened at low temperatures, as
expected from the vanishing of the intermediate fixed point.
If instead $V=0.1D$, there is a finite critical coupling,
$J_c\approx 1.3D$; now an exchange $J=1.5D$ ensures that the impurity
is screened for $T\ll T'_X\approx 10^{-3}D$.

In order to investigate the sensitivity of the preceding results to the precise
form of Eq.~(\ref{puredos}), the effect of three changes to the density of
states will be briefly examined:

(i) In real systems, the power-law variation of $\rho(\epsilon)$ is
restricted to an energy range $|\epsilon|\le\Delta$, with
$\rho(\epsilon)\approx\rho(\Delta)$ for $\Delta<|\epsilon|\le D$.
At temperatures $T\gg\Delta$, one finds the standard Kondo physics of
metallic systems.
For $T\ll\Delta$, however, the impurity ``sees'' a pure power-law density of
states, so the results above still apply, albeit with $J$ replaced by
$J_{\text{eff}}(\Delta)>J$.
This enhancement makes realization of the Kondo effect more plausible
for systems with $r\gtrsim 1$, a range in which the critical couplings would
otherwise be unphysically large.
Figure~\ref{fig:chi2} illustrates the point:
with $\Delta=10^{-3}D$ and $V=0.3D$, a coupling $J=0.5D$ fully screens
the impurity, for both $r=1$ and $2$.

(ii) Band asymmetry can be introduced by shifting the Fermi level away
from the band center, while retaining the power-law variation of
$\rho(\epsilon)$ about the new Fermi energy.
Although this changes the $\xi_{M-1}$'s entering Eq.~(\ref{H_M}),
and generates nonzero coefficients $\varepsilon_M$, the physical effects are
little different from those of potential scattering.

(iii) Partial filling of the pseudo-gap may introduce a lower cutoff $\Delta'$
on the power-law density of states, with $\rho(\epsilon)\approx\rho(\Delta')$
for $|\epsilon|<\Delta'$.
The finite value of $\rho(0)$ must eventually produce a standard, metallic
Kondo effect.
However, for $r\gtrsim 1$ and most plausible values of $J$, $\Delta$ and
$\Delta'$, this occurs at inaccessibly low temperatures.

{\em Discussion:\/}
Although large-$N$ studies of gapless systems
\cite{Withoff,Cassanello,Borkowski}
have treated symmetric bands with zero potential scattering,
they have found no sign of the Kondo effect disappearing for any power $r$.
This discrepancy with Ref.~\cite{Chen} and the present work may stem from the
mean-field nature of the large-$N$ method, or from the
symmetry-breaking that is implicit, for all $N>2$, in the restriction
that the impurity level be singly occupied.

The $r=\infty$ limit of the restricted density of states defined in (i) above
describes an insulator with gap $2\Delta$.
The results of this work are entirely consistent with those known for
gapped systems.
At particle-hole symmetry, an impurity in an insulator retains its moment,
no matter how large $J$ is made;
away from this symmetry, the spin is screened provided that
$J>J_c\approx 2D/\ln(D/\Delta)$ \cite{Takegahara}.
The mapping of an impurity in an $s$-wave BCS superconductor onto
Eq.~(\ref{H1D}) necessarily introduces particle-hole asymmetry, resulting
in a finite $J_c$ \cite{Satori}.
Figure~\ref{fig:chi2} demonstrates the similarities between the impurity
susceptibilities for $r=1$, $2$ and $\infty$.
However, the low-temperature decrease in the impurity resistivity is a
signature of gapless systems which has no counterpart in insulators.

This study shows that the behavior of dilute spin impurities in gapless
Fermi systems differs qualitatively from that in metals or insulators.
In most gapless materials, a sufficiently large exchange coupling, $J>J_c$,
results in complete screening of the impurity, marked by the vanishing of
$T\chi_{\text{imp}}$ and a low-temperature {\em decrease\/} in the
impurity resistivity; for $J<J_c$, the impurity asymptotically decouples
from the band.
Systems which lie close to particle-hole symmetry are unlikely to exhibit
a Kondo effect because in such cases $J_c$ becomes unphysically large.
Further work is needed to extend these results to degeneracies $N>2$,
to obtain the full temperature dependence of the resistivity,
and to treat self-consistently magnetic impurities in unconventional
superconductors.

I thank P.\ Hirschfeld and A.\ Schiller for many helpful discussions;
C.\ Cassanello, E.\ Fradkin, and K.\ Ziegler for informative conversations;
L.\ Oliveira for pointing out the form of Eq.~(\ref{xi_M});
and T.\ Costi for drawing my attention to Ref.~\onlinecite{Chen}.
This work was supported in part by NSF Grant No.\ DMR--9316587.


\begin{thebibliography}{18}
\bibitem{units}
    The units are such that $k_B=g\mu_B=1$, where $g$ is the
    electronic $g$-factor and $\mu_B$ is the Bohr magneton.
\end{thebibliography}
\end{document}